# Intertwined density waves in a metallic nickelate

Junjie Zhang[1,2]*, D. Phelan[1], A. S. Botana,[3] Yu-Sheng Chen[4], Hong Zheng[1], M. Krogstad[1], Suyin Grass Wang[4], Yiming Qiu[5], J. A. Rodriguez-Rivera[5,6], R. Osborn[1], S. Rosenkranz[1], M. R. Norman[1] and J. F. Mitchell[1]*

[1]Materials Science Division, Argonne National Laboratory, Lemont, Illinois 60439, United States.
[2]Institute of Crystal Materials, Shandong University, Jinan, Shandong 250100, China.
[3]Department of Physics, Arizona State University, Tempe, Arizona 85287, United States.
[4]ChemMatCARS, The University of Chicago, Lemont, Illinois 60439, United States.
[5]NIST Center for Neutron Research, National Institute of Standards and Technology, Gaithersburg, Maryland 20899, United States.
[6]Department of Materials Science, University of Maryland, College Park, Maryland 20742, United States.

*e-mail: junjie@sdu.edu.cn; mitchell@anl.gov

## Abstract

Nickelates are a rich class of materials, ranging from insulating magnets to superconductors. But for stoichiometric materials, insulating behavior is the norm, as for most late transition metal oxides. Notable exceptions are the 3D perovskite $LaNiO_3$, an unconventional paramagnetic metal, and the layered Ruddlesden-Popper phases $R_4Ni_3O_{10}$, (R=La, Pr, Nd). The latter are particularly intriguing because they exhibit an unusual metal-to-metal transition. Here, we demonstrate that this transition results from an incommensurate density wave with both charge and magnetic character that lies intermediate in behavior between the metallic density wave seen in chromium metal and the insulating stripes typically found in layered nickelates. As such, $R_4Ni_3O_{10}$, which appears to be the first known example of an itinerant spin density wave in a $3d$ transition metal oxide, represents an important bridge between the paramagnetism of 3D metallic $LaNiO_3$ at higher nickel valence and the polaronic behavior of quasi-2D $R_{2-x}Sr_xNiO_4$ at lower nickel valence.

## Introduction

The central challenge in harnessing the unrivaled diversity of transition metal oxides (TMO) is understanding (and manipulating) how charge and spin influence their properties. Gaining such understanding and control is made difficult by the many highly correlated degrees of freedom ubiquitously present in TMO. As a case in point, the family of rare earth nickel oxide perovskites, $RNiO_3$ (R=La, Pr-Lu), presents an important archetype TMO for investigating the nexus between charge localization and itinerancy and their correlation with magnetism. The well-studied phase diagram of these perovskites shows that for rare earth $R^{3+}$ ions with radius smaller than $Nd^{3+}$, a metal-insulator transition (MIT) occurs coincidently with a structural phase transition at $T_S$, which itself lies at a temperature higher than a subsequent antiferromagnetic transition, temperature, $T_N$. For R=Nd and Pr, $T_S = T_N$[1]. On the other hand, the La member of this series shows none of this phenomenology. $LaNiO_3$ is a good metal that remains rhombohedral (space group $R\bar{3}c$) at all temperatures, and it shows no evidence of intrinsic, long-range ordered antiferromagnetism in bulk samples[2-4]. Nonetheless, evidence of short-range bond disproportionation has been reported[5], and recently $LaNiO_3$ has been argued to lie proximate to an antiferromagnetic quantum critical point (QCP) in clean epitaxial thin films[6]. The possibility of a QCP associated with a $T=0$ MIT has also been inferred by tunneling measurements through the appearance of a pseudogap in the electronic density of states[7]. Thus, the $LaNiO_3$ system lies poised between localized and itinerant behavior, between magnetic order and disorder, and at the crossroads between discontinuous and continuous phase transitions.

In this spirit, the quasi-2D phases, $R_4Ni_3O_{10}$ (R=La, Pr, Nd), the $n=3$ member of the Ruddlesden-Popper (R-P) series $R_{n+1}Ni_nO_{3n+1}$, can serve as proxies for revealing ordering phenomena that are latent in the perovskites. As has been shown previously[8-13], members of this series undergo a metal-to-metal transition (MMT) near 150 K, with $T_{MMT}$ depending on R. The precise mechanism of this MMT has remained an open question for nearly 25 years[8,10]. Here we report that the MMT in $La_4Ni_3O_{10}$ results from intertwined charge-density (CDW) and spin-density (SDW) waves. Combining single crystal synchrotron x-ray and neutron diffraction, we find charge and spin-superlattice (SL) reflections below the MMT with propagation vectors $q_c = (0,q_c,0)$ and $q_s = (0,1-q_s,0)$, respectively, with $q_c=2q_s$ as expected for a system with coupled charge and spin order. We present models for these intertwined density waves that reproduce the experimental observations semi-quantitatively. In the charge sector, our data are uniquely consistent with a CDW centered on the Ni sites, establishing that the low energy electronic degrees of freedom in this system lie on the transition metal rather than on oxygen. The lack of local moments on La unequivocally establishes that the SDW is associated with the Ni ions as well, meaning that the $3d$ oxide $La_4Ni_3O_{10}$ joins a relatively small family of itinerant SDW systems that includes chromium[14], $Sr_3Ru_2O_7$[15] (in a magnetic field), organic conductors[16], hole-doped 122 iron pnictides[17], and several heavy fermion systems[18,19]. To the best of our knowledge, $La_4Ni_3O_{10}$ is the first verified example of a *3d* oxide itinerant SDW material.

## Results

**Physical properties.** Figs. 1a-c show resistivity, magnetic susceptibility, and heat capacity measured on $La_4Ni_3O_{10}$ single crystals, which adopt a pseudo-orthorhombic *Bmab* structure shown in Fig. 2a (properly indexed as monoclinic, $P2_1/a$, see Supporting Information, Section 1). The in-plane resistivity (Fig. 1a) drops with decreasing temperature, indicating metallic behavior, and an anomaly is observed at $T_{MMT} \approx 140$ K. We note that the temperature-dependent transport behavior of La$_4$Ni$_3$O$_{10}$ near $T_{MMT}$ resembles that reported for several density wave materials including chromium[14], purple bronzes[20], and rare-earth tritellurides[21]. The magnetic susceptibility (Fig. 1b) shows a sharp decrease at $T_{MMT}$, but it does not follow either a Pauli or Curie-Weiss form above $T_{MMT}$ in the temperature range measured, as it monotonically increases with increasing temperature. It does, however, resemble that of CDW materials such as the one-dimensional K$_{0.3}$MoO$_3$[20], in which the temperature dependence of the susceptibility above the MMT was attributed to CDW fluctuations and a pseudogap in the electronic density of states[22]. The heat capacity of La$_4$Ni$_3$O$_{10}$ at low temperature (inset, Fig. 1c) was fit to the typical form, $C_p/T=\gamma+\beta T^2$, where $\gamma$ is the Sommerfeld coefficient and the $\beta T^2$-term arises from phonons. The fit leads to $\gamma =13.3$ mJ mole$^{-1}$ K$^{-2}$ and $\beta=0.37$ mJ mole$^{-1}$ K$^{-4}$, which agrees with a previous measurement of $C_p$ on polycrystalline samples by Wu et al.[23] ($\gamma =13.5$ mJ mole$^{-1}$ K$^{-2}$) and by Kumar et al.[9] ($\gamma =15.5$ mJ mole$^{-1}$ K$^{-2}$); however, we find a significantly larger Debye temperature ($\theta_D = 450$ K) than that reported by either group ($\theta_D = 256$ K[23], 384 K[9], respectively). For comparison, reported values for single crystal LaNiO$_3$ are $\gamma = 11.7$ mJ mole$^{-1}$ K$^{-2}$ and $\theta_D = 465$ K[2]. From the Sommerfeld coefficient, $\gamma=\pi^2 k_B^2 N(E_F)/3$, the density of states at the Fermi level $N(E_F)$ in La$_4$Ni$_3$O$_{10}$ is estimated to be 1.9 states eV$^{-1}$ per Ni, significantly less than that of LaNiO$_3$, 5.0 states eV$^{-1}$ per Ni.

**Charge density wave order.** Using single crystal synchrotron x-ray diffraction, we observed charge superlattice (SL) peaks in La$_4$Ni$_3$O$_{10}$ below the MMT, indicating the appearance of a long-range charge-ordered state. The temperature dependence of the integrated intensity of one such SL reflection is shown in Fig. 1d. Its onset at $T_{MMT}$ correlates with resistivity, magnetic susceptibility and heat capacity measurements. We note that these charge SL reflections have intensities typically 10$^4$ times weaker than the fundamental Bragg reflections, explaining why they have been unobserved in prior studies of polycrystalline samples.

Figs. 2b and 2c show the $hk0$ plane in reciprocal space at 200 K and 100 K, respectively. At 200 K, strong fundamentals appearing at integers $hk0$ obey the $Bmab$ reflection condition that they are both even, but additional scattering attributable to the $P2_1/a$ cell is weakly observed as peaks that violate this rule. As shown in Fig. 2c, sharp first order superlattice reflections appear between certain fundamentals at 100 K, located at $h=2n+1$ and $k=2m+1 \pm q_c$, ($m$, $n$ integers). The appearance of SLs along $k$ coincides with the observation of a pronounced anomaly in the lattice parameter $b$ at the MMT[8,9]. In this particular crystal, the SL reflections appear only along $k$, indicating a single domain modulation despite the near metrical equivalence of $a$ and $b$, which differ by <1%. We note that several other specimens that we investigated did show the presence of orthogonal modulations expected of a multidomain state. Higher harmonics of the SL are not observed, demonstrating a sinusoidal character to the CDW. This tends to rule out alternatives such as real-space charge stripes like those found in the ⅓-hole doped single layer R-P phase La$_{1.67}$Sr$_{0.33}$NiO$_4$[24] or square planar trilayer La$_4$Ni$_3$O$_8$[25], although the possibility of a locally commensurate stripe phase with $q_c$=⅔ and a phase slip every $\approx$11 diagonal rows cannot be ruled out[26].

Line cuts along (-1, $k$, 0) and (0, $k$, 19) are shown in Figs. 2d and 2e, respectively. By fitting the peaks, a modulation wavevector $\boldsymbol{q_c}$=0.76$\boldsymbol{b}$* was obtained (see also Figs. S1 and S2). The width of the SL peaks is somewhat broader than that of the corresponding fundamental peaks (Fig. S3), indicating long-range charge correlations (in-plane correlation length, $\xi_{ab}\sim$100 Å). This long-range correlation of charge order is comparable to that reported for stripe-order in La$_{1.67}$Sr$_{0.33}$NiO$_4$[24] and charge order in YBa$_2$Cu$_3$O$_{6.67}$[27].

Figs. 2f and 2g show $\bar{1}kl$ and $\bar{2}kl$ planes at 100 K, and cuts along $l$ are shown in Figs. 2h-k. SL peaks along $l$ occur at integers obeying the selection rule that $h+l$ is odd. This selection rule is shown in Figs. 2h,j. For $h=2n+1$, SL peaks are located at $l=2m$, while for $h=2n$, SL peaks are observed at $l=2m+1$ ($m$, $n$ integers). In contrast, when $h$ and $l$ have the same parity, no SL reflections are observed above background. SL peaks in these figures are observed for $h,k$ of the same parity ($k$ is defined by the integer from which $q_c= 0.76b^*$ is measured). As shown in Figs. 2i,k weaker SL peaks are also found for $h,k$ of mixed parity. Another notable feature is that the intensity in these cuts along $c^*$ is largest at $l=0$ and 14 for SL peaks associated with odd $h$ and odd $k$ fundamentals (Fig. 2h), but at $l=7$ for SL peaks at even $h$ and even $k$ (Fig. 2j). The interval between maxima is similar to the ratio of the length of the $c$ axis to the Ni-O layer separation within a trilayer ($\tau = c/d_{\text{Ni-Ni}} = 7.2$). The intensity distribution along $c^*$ thus reflects a sinusoidal modulation of the trilayer unit structure factor with period $\tau$. A model of the CDW discussed below captures these features of the CDW diffraction pattern. Finally, the width of SL peaks along $l$ is significantly broader than that of the nearby Bragg peaks, indicating a finite correlation length, $\xi_c$, of the charge order. Analysis of the peak width yields $\xi_c \approx 0.7c \approx 21$Å (Fig. S4), verifying weak correlation of the charge order between neighboring unit cells along $c$ required to generate the selection rules discussed above.

**Spin density wave order.** Single crystal neutron diffraction data were measured to test for the presence of an SDW concomitant with the CDW. If such an SDW were present, a coupling between these two density waves is expected to occur such that $q_c = 2q_s$[28]. Fig. 3a shows the 0$kl$ plane in the reciprocal space of the neutron scattering data, with a background measured at 180 K subtracted (additional measurements in the $hk$0 plane are discussed in the SI, Fig. S5). SL reflections attributed to magnetic scattering are observed at $k=1\pm q_s$, $-1\pm q_s$. A cut along (0, $k$, 2) is presented in Fig. 3b, where magnetic SL reflections appear at $k= -1.38$ and $-0.62$, i.e., $-1\pm q_s$. Noting that (0, 1) corresponds to the "($\pi$, $\pi$)" antiferromagnetic wavevector that is frequently employed in the cuprate literature (i.e, with reference to an undistorted 3.8 Å × 3.8 Å tetragonal sub-cell), the SDW propagation vector in La$_4$Ni$_3$O$_{10}$ was determined to be $q_s=0.62b^*$, that is $q_s=(0,1-q_s,0)$, with $q_s=0.38$. With this assignment of the magnetic SL, $q_c = 2q_s$ (Fig. 3d), as expected for a coupling between the CDW and SDW. The cut along (0, 1-$q_s$, $l$) shown in Fig. 3c evidences strong SL peaks along $c^*$ at $l= \pm 2, \pm 6, -10$, i.e., $l = 4n+2$. This intensity distribution suggests antiferromagnetically coupled planes separated by $nc/8$, with $n$ odd. This would occur if the inner planes were non-magnetic and the outer planes were $\pi$ out of phase. However, a closer inspection shows weaker intensity appearing at $l=\pm 4$ and $-8$ as well, reflecting that the outer plane spacing is in reality $c/3.6$ instead of $c/4$.

We show in Fig. 1e the temperature dependence of the integrated intensity measured at (-1,0.38,0). As expected, the onset of intensity is observed at $T_{MMT}$, coincident with the CDW. There is no compelling evidence in Figs. 1d,e for assigning charge or spin as the primary order parameter, and the onset temperature is the same for both within experimental resolution. In these ways, La$_4$Ni$_3$O$_{10}$ behaves similarly to the related La$_4$Ni$_3$O$_8$ system containing square planar Ni[29].

**Density wave models and simulations.** We begin with the SDW. Prominent peaks at $l=2$ and 6 along the (0, 0.62, $l$) cut (Fig. 3c) indicate an approximate pattern of $l=4n+2$ when $h=0$. This selection rule implies that the magnetic stacking pattern of the six planes in the unit cell (three per trilayer) is ↑, −, ↓; ↑, −, ↓, where − represents a node. This argument holds regardless of the direction of spin polarization. Had the planes been spaced precisely $d=c/8$ apart (as in the related compound La$_4$Ni$_3$O$_8$[29]), then the $l=4n+2$ selection rule would be exact. We note that such a magnetic ground state, with an SDW node on the inner planes, is very unusual. A related example may be layered cuprates, where the inner and outer planes differ electronically due to different effective doping levels[30]. In La$_4$Ni$_3$O$_8$, the magnetic stacking pattern is ↑,↓,↑ instead (leading to a prominent peak at $l=4$, which is notably absent here), with ↑, −, ↓ being an excited state.

The simplest model is to assume the in-plane behavior of the SDW has the form $\cos(\boldsymbol{q}_\perp \cdot \boldsymbol{r}_\perp)$ with $\boldsymbol{q}_\perp=(0,1-q_s)$ and $\boldsymbol{r}_\perp$ the in-plane coordinates of the Ni ions. Our calculations implicitly assume that the spin direction is along $\boldsymbol{a}$ (i.e., in the basal plane and perpendicular to $\boldsymbol{q}_s$), which is consistent with our experimental observations[31], although additional (polarized) measurements would be required for absolute determination of the spin direction. For a given plane, the SDW corresponds to diagonal rows of aligned spins as found in La$_4$Ni$_3$O$_8$[29] and La$_{1.67}$Sr$_{0.33}$NiO$_4$[28]. We choose a commensurate value of $q_s=3/8$ to facilitate comparison to real space models. The resulting SDW pattern is shown in Fig. 4a. The intensity distribution is calculated from $I(\boldsymbol{k}) = |\sum_{\boldsymbol{r}} c_z e^{i\boldsymbol{k}\cdot\boldsymbol{r}} \cos(\boldsymbol{q}_\perp \cdot \boldsymbol{r}_\perp)|^2$, with $c_z$=1,0,-1;1,0,-1 encoding the magnetic stacking pattern along $c$ discussed above with the sum $\boldsymbol{r}$ over Ni atoms (normalized by their number). The resulting (0, $k$, 2) and (0, 5/8, $l$) cuts are shown in Figs. 4c and 4d, respectively, and should be compared to the experimental data in Figs. 3b and 3c, respectively. In Fig. 4d, note the small peak at $l=4$ and the larger peaks for even values of $l$ beyond $l=6$. This arises from the fact that the planes are spaced within the trilayer by $\sim c/7.2$ rather than by $c/8$. In Fig. S7, results for different assumed SDW stackings along $c$ are also shown, along with a comparison of cuts along $l$ for $h=0$ and $h=-1$ in Fig. S8.

We now turn to the CDW peaks. Strong CDW peaks are seen for $h+l=2n+1$ and $k+l=2n+1$. This pattern, as well as the presence of a strong peak at $l=7$ for certain momentum cuts like (0,-3.24,$l$), implies that the CDW is (1) present on all planes, (2) is in-phase within a trilayer, and (3) is out-of-phase between successive trilayers. For (1), because $d\sim c/7.2$ (explaining the strong peak at $l=7$) and because the pattern of CDW $l$ harmonics does not reflect that of the SDW discussed above, the CDW must be present on all planes. Condition (2) derives from the $Bmab$ (1/2,0,1/2) face-centered translation reflection condition, i.e., $h+l=2n$. This general $Bmab$ condition becomes $h+l=2n+1$ if the modulation in successive trilayers is $\pi$ out of phase. The other condition, $k+l=2n+1$, occurs because the Ni ion positions deviate only slightly from (00$z$). This slight deviation generates weak reflections that satisfy $k+l=2n$ as well. The resulting CDW pattern, modeled by $\boldsymbol{q}_\perp=(0,2-q_c)$ with $q_c=3/4$ (and $c_z$=1,1,1;-1,-1,-1), is shown in Fig. 4b. Representative cuts demonstrating strong CDW peaks are shown in Figs. 4e and 4f. Note that $\boldsymbol{q}_c = 2\boldsymbol{q}_s$, as expected when relating the CDW peaks to those of the SDW. Results for different assumed CDW stackings along $c$, as well as contrasting different $k$ and $l$ cuts, are shown in Figs. S9-S11.

The fact that a simple cosine wave can explain the data, along with the observed incommensurability and the additional observation that this is a metal to metal transition, all support the picture of an itinerant density wave, as seen in chromium[14]. Unlike chromium, however, the CDW peaks in La$_4$Ni$_3$O$_{10}$ are quite strong, and there is no evidence (Figs. 1d,e) that the CDW amplitude is secondary (i.e., quadratic in the SDW amplitude). That is, $La_4Ni_3O_{10}$ is more reminiscent of its reduced analog, $La_4Ni_3O_8$, though in that case, one has a semiconductor to insulator transition and a real space charge- and spin-stripe scenario[25,29]. In some sense, one could regard $La_4Ni_3O_{10}$ as a doped version of $LaNiO_3$, which is a metal known to lie near an antiferromagnetic quantum critical point. Indeed, $La_4Ni_3O_{10}$, with an average Ni valence of 2.67 ($d^{7.33}$), corresponds to $Ni^{2+}$ and $Ni^{3+}$ in a 1:2 ratio and as such is ⅓ electron-doped relative to the parent $LaNiO_3$ perovskite.

**Discussion**

Recent reports on the electronic structure of the related trilayer R-P phases $Pr_4Ni_3O_{10}$[12] and $Nd_4Ni_3O_{10}$[10] have stressed the importance of differentiating the inner layer and outer layers of the trilayer blocks. Indeed, because of symmetry, some degree of charge differentiation between the layers is inevitable. The model presented above for the CDW posits a uniform charge density modulation in all three layers, but other models we have studied with nonuniform weights only differ in a quantitative sense (Fig. S9). We note that differential doping could offer an explanation for the node in the SDW observed on the inner layer, as suggested for multi-layered cuprates[30], though we remark that our modeling unambiguously finds that the CDW is present on the inner plane as well.

The effect of chemical or physical pressure on the MMT in the $R_4Ni_3O_{10}$ (R=La, Pr, Nd) series has recently been shown through the systematic response of $T_{MMT}$ to these parameters[10,11]. The monotonic variation of the transition is consistent with a Fermi surface driven instability sensitive to variables that can modify the electronic band structure and hence the relevant nesting vector(s)[32,33]. Indeed, we find that by substituting Pr for La that the propagation vector responds to the effect of chemical pressure, albeit weakly, with $q_c = 0.78$ for $Pr_4Ni_3O_{10}$, and $q_c = 2q_s$ (Fig. S6). In support of this nesting picture, we show calculations of the susceptibility from band theory in the SI (Fig. S13).

A distinguishing feature of $La_4Ni_3O_{10}$ is that not only do the SDW and CDW onset at the same temperature, but also they both appear to be primary order parameters in the sense of the Landau theory of phase transitions (Figs. 1d,e). Notably, $La_4Ni_3O_{10}$ shares this behavior with its reduced, oxygen deficient analog, $La_4Ni_3O_8$[29], although the latter is insulating below the phase transition. This contrasts with chromium, where the CDW/strain wave is a secondary order parameter[34].

With its trilayer structure, the ground state of quasi-2D $La_4Ni_3O_{10}$ lies at a crossover between the paramagnetic 3D metal $LaNiO_3$ and the insulating, more 2D single layer nickelates, $La_{2-x}Sr_xNiO_4$ ($x<1$). The implication is that this intermediate behavior reflects an enhanced coupling among charge, spin, and lattice degrees of freedom vis-à-vis the perovskite, but falls short of that found in the single-layer materials. Rather than the static real-space charge and spin stripes found in the latter materials when $x<0.5$[28], the result of this moderately strengthened coupling in $La_4Ni_3O_{10}$ is an incommensurate charge- and spin-density wave order that we believe is the first to be verified in a *3d* oxide. Metallic conductivity coexisting with the CDW and SDW implies a partial gapping of the Fermi surface at $T_{MMT}$. Indirect support for this latter point comes from the specific heat, where $N(E_F)$ for metallic $LaNiO_3$ is nearly three times larger than that of $La_4Ni_3O_{10}$. Future studies using, for instance, high resolution ARPES and tunneling could shed light on this putative gapping[35].

Finally, the relation of the incommensurate modulations observed here and the antiferromagnetic QCP identified for the perovskite LaNiO$_3$ is not clear. This could be studied, for instance, by hole-doping La$_4$Ni$_3$O$_{10}$ toward metallic LaNiO$_3$. In a like vein, the relation of La$_4$Ni$_3$O$_{10}$ to insulating La$_{2-x}$Sr$_x$NiO$_4$ could be investigated by electron-doping instead, or else by studying the bilayer homolog La$_3$Ni$_2$O$_7$. From such studies, it should be possible to build a unified understanding of this exceptional group of quantum materials.

**Methods**

**Sample growth and characterization.** Single crystals of La$_4$Ni$_3$O$_{10}$ were grown using the high-pressure floating zone furnace at 20 bar O$_2$ (Model HKZ, SciDre, Germany[36]). The sample growth and characterization methods, including oxygen stoichiometry, have been reported previously[8].

**Magnetic Susceptibility**. Magnetic susceptibility measurements were performed on single crystals using a Quantum Design MPMS3 SQUID magnetometer[36]. Specimens were attached to a quartz holder using a minute amount of glue. ZFC-W (Zero-field cooling with data collected on warming), FC-C (field cooling with data collected on cooling) and FC-W (field cooling with data collected on warming) data with magnetic field **H**∥*ab* and **H**⊥*ab* were collected between 1.8 and 300 K under an external field of 0.4 T. In the ZFC-W protocol the sample was cooled in zero field to 10 K at a rate of 35 K/min and then to 1.8 K at a rate of 2 K/min, and DC magnetization recorded on warming (2 K/min). In the FC-C and FC-W protocols, the magnetization was recorded (2 K/min) in a fixed field of 0.4 T.

**Electrical resistivity**. Resistivity was measured using a four-terminal method with contacts made by depositing gold pads. Temperature was controlled using the Quantum Design PPMS in the temperature range of 1.8−300 K.

**Heat capacity**. Heat capacity measurements were performed on a Quantum Design PPMS[36] in the temperature range of 1.8-300 K. Apiezon-N[36] vacuum grease was employed to fix crystals to the sapphire sample platform. The specific heat contribution from sample holder platform and grease was determined before mounting sample and subtracted from the total heat capacity.

**X-ray diffuse scattering**. Synchrotron X-ray single crystal diffraction measurements were performed at beamline 15-ID-D at ChemMatCARS (University of Chicago) at the Advanced Photon Source, Argonne National Laboratory. Data were collected with a 1M Pilatus area detector using synchrotron radiation (λ=0.41328 Å) at 100, 110, 120, 125, 127, 129, 131, 133, 135, 137, 139, 145, 200 K with temperature controlled by flowing nitrogen gas. To cover a sufficient volume of reciprocal space, ω and φ scans were used, with 1800 frames (0.2 deg/frame, continuous scan) collected at each ω angle. Several single crystals were checked in this experiment, and the SL peaks are reproducible. The 3D reciprocal space volumes were generated from the data using the CCTW package[37] and visualized using NeXpy[38], which was used to produce the linecuts. To obtain the precise wavevector, data were collected at beamline 33-BM-C[39] (Advanced Photon Source) using a point detector. The wavevector $q_c$=0.76$b$* was obtained by fitting the peaks.

**Neutron elastic scattering.** Unpolarized measurements in the *hk*0 and 0*kl* scattering planes were performed on the MACS cold neutron multi-axis spectrometer at the NIST Center for Neutron Research (NCNR) with λ=4.05 Å. Four crystals were coaligned for the *hk*0, and a crystal of mass ≈40 mg was used for 0*kl* plane, and data were collected at 1.6 and 180 K. The SDW order parameter was determined by following the evolution of the intensity at (-1,0.38,0) on warming.

**Density wave models and simulations.** For numerical purposes, the intensities were calculated using 10 unit cells along *c* (20 trilayers) and $25^2$ orthorhombic unit cells in the plane (for a total of 75000 Ni atoms). The slight *Bmab* deviations of the outer plane Ni ions from their *I4/mmm* positions were taken into account, using atomic coordinates given in Ref.8 Details are given in the SI, Section 6 (Figs. S7-11) along with real space stripe simulations in the SI, Section 7 (Fig. S12).

**Computational details.** We performed first-principles calculations using the WIEN2k[40] code with the Perdew-Burke-Ernzerhof[41] version of the generalized gradient approximation. A Fourier series spline fit[42] to the bands was made with 2736 face centered orthorhombic Fourier functions fit to 448 *k* points in the irreducible wedge of the Brillouin zone. Both the density of states and the susceptibility were calculated using a tetrahedron decomposition of the Brillouin zone[43] ($3 \times 8^n$ tetrahedra in the irreducible wedge with n=6 used for the density of states and n=5 for the susceptibility). These results are presented in the SI, Section 9 (Fig. S13).

**Data availability.** The data that support the findings of this study are available within the article or from the corresponding author upon request.

**Acknowledgements**

This work was supported by the US Department of Energy, Office of Science, Basic Energy Sciences, Materials Science and Engineering Division. The work at Shandong University was supported by the Qilu Young Scholar Program of Shandong University, and the Taishan Scholar Program of Shandong Province. Use of the Advanced Photon Source at Argonne National Laboratory was supported by the U.S. Department of Energy, Office of Science, Office of Basic Energy Sciences, under Contract No. DE-AC02-06CH11357. ChemMatCARS Sector 15 is supported by the National Science Foundation under grant number NSF/CHE-1836674. Access to MACS was provided by the Center for High Resolution Neutron Scattering, a partnership between the National Institute of Standards and Technology and the National Science Foundation under Agreement No. DMR-1508249. The authors thank Drs. J. W. Freeland, W. E. Pickett and Vladimir A. Stoica for helpful discussions. J.Z. thanks Dr. Evguenia Karapetrova for her help with the measurements on $R_4Ni_3O_{10}$ (R=La, Pr) at Beamline 33-BM-C, Advanced Photon Source.

**Author contributions**

J.Z., J.F.M. and D.P. conceived and designed the experiments. J.Z. grew single crystals with help from H.Z. J.Z. performed synchrotron x-ray single crystal diffraction with the help of Y.S.C. and S.G.W. J.Z., S.R. and D.P. performed neutron diffraction with the help of Y.Q. and J.A.R.-R. M.R.N. developed the charge and spin density wave models and performed simulations of the diffraction patterns. A.S.B. and M.R.N. performed DFT calculations and electronic susceptibility calculations. J.Z., D.P., M.K., J.F.M., M.R.N., S.R. and R.O. analyzed data. J.Z., J.F.M., M.R.N., and D.P. wrote the manuscript with contributions from all authors.

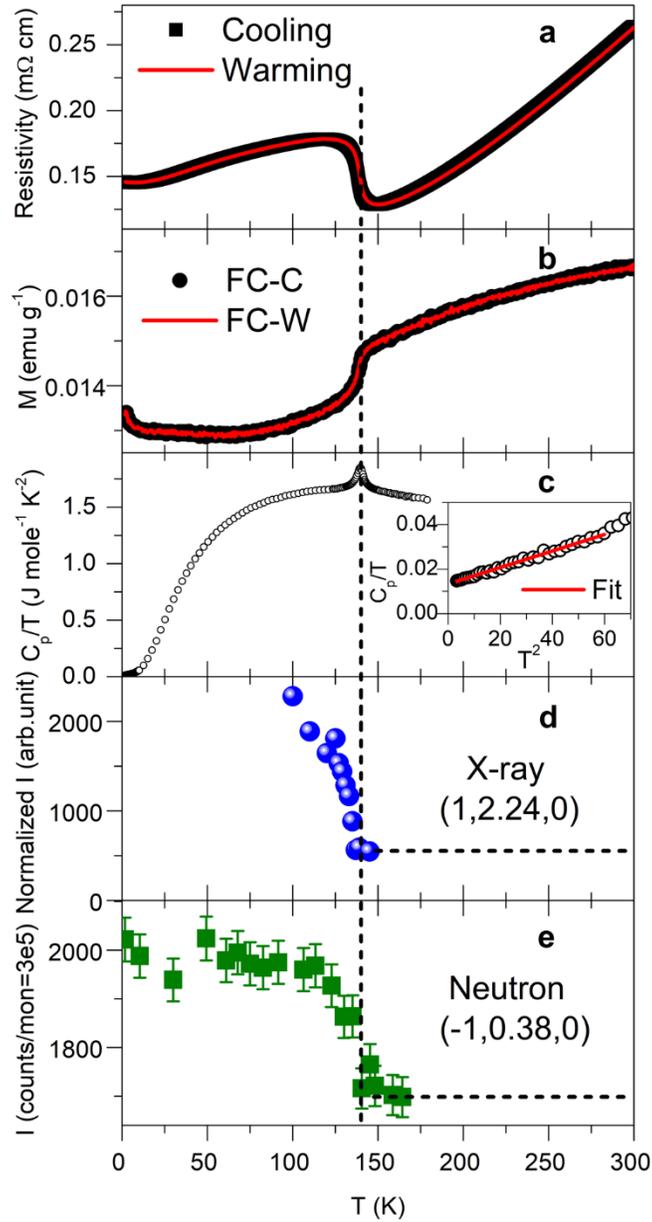

**Fig. 1 | Physical properties and order parameter of La$_4$Ni$_3$O$_{10}$. a,** Resistivity. **b,** Magnetic susceptibility measured under 0.4 T upon cooling and warming. **c**, Heat capacity. Inset show fit for low temperature data. **d**, Temperature dependence of integrated intensity of (1, 2.24, 0) measured with x-rays. **e**, Temperature dependence of integrated intensity at (-1, 0.38, 0) with measured with neutrons. Error bars represent one standard deviation.

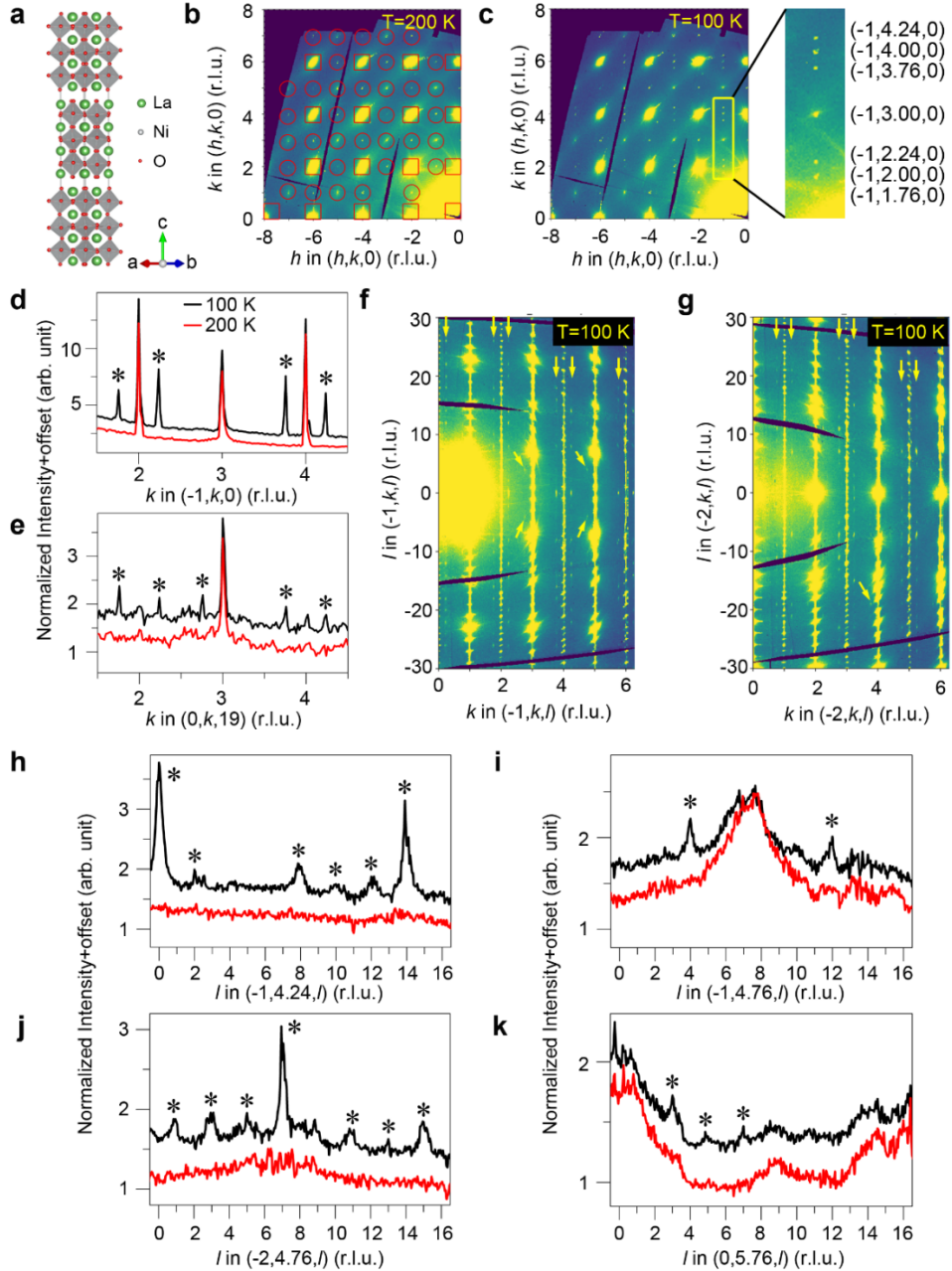

**Fig. 2 | Charge density wave order in La$_4$Ni$_3$O$_{10}$. a**, Crystal structure of La$_4$Ni$_3$O$_{10}$. The unit cell contains two trilayer perovskite-like blocks that are related through the *B*-centering operation of the *Bmab* space group. **b**, *hk*0 plane at 200 K as measured at Sector 15-ID-D. The data is integrated over 0.05 r.l.u. in *l*. Hollow squares, fundamental Bragg peaks from high symmetry *Bmab*; hollow circles, Bragg peaks resulting from monoclinic distortion to *P*2$_1$/*a*. **c**, *hk*0 plane at 100 K. **d-e**, Line cuts through data along *k* direction at 100 and 200 K. The cuts have been integrated over 0.04 r.l.u. in *h* and 0.05 r.l.u. in *l*. **f**, $\bar{1}kl$ plane at 100 K. The data are integrated over 0.02 r.l.u. in *h* (*h*=-1). **g**, $\bar{2}kl$ plane at 100 K. **h-k**, Line cuts through data along *l* direction at 100 and 200 K. Note red for 200 K and black for 100 K. The cuts have been integrated over 0.02 r.l.u. in *h* and 0.04 r.l.u. in *k*. In (**d, e, h-k**), the data are shifted for clarity. Asterisks (*) mark the superlattice reflections. Yellow arrows in (**b, c, f, g**) point to superlattice reflections.

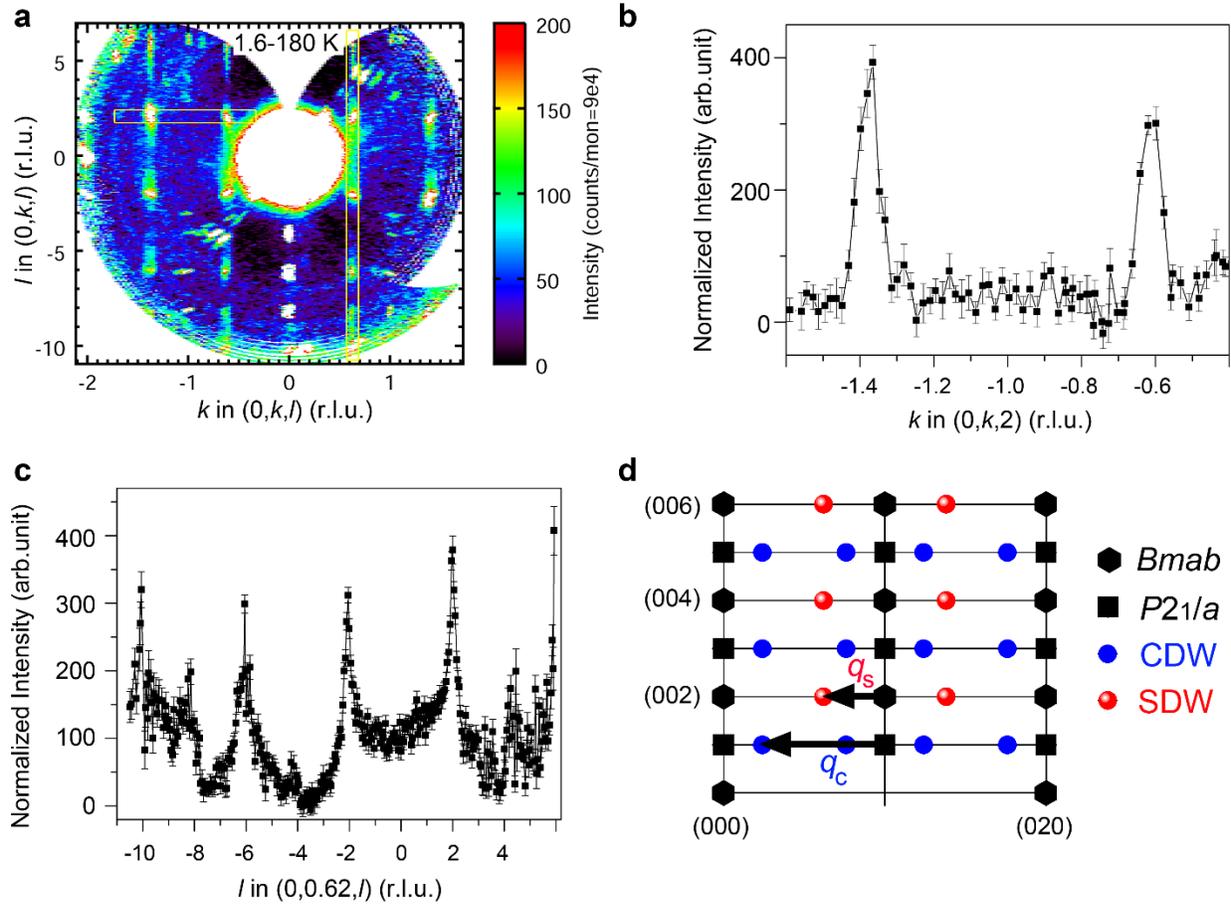

**Fig. 3 | Spin density wave order in La$_4$Ni$_3$O$_{10}$. a**, 0$kl$ plane with background (180 K) subtracted from 1.6 K data. **b-c**, Line cuts along (0, $k$, 2) and (0, 0.62, $l$), respectively. **d**, Schematic of the 0$kl$ plane showing location of *Bmab* and *P*2$_1$/*a* fundamentals and both CDW and SDW superlattice reflections. Error bars represent one standard deviation.

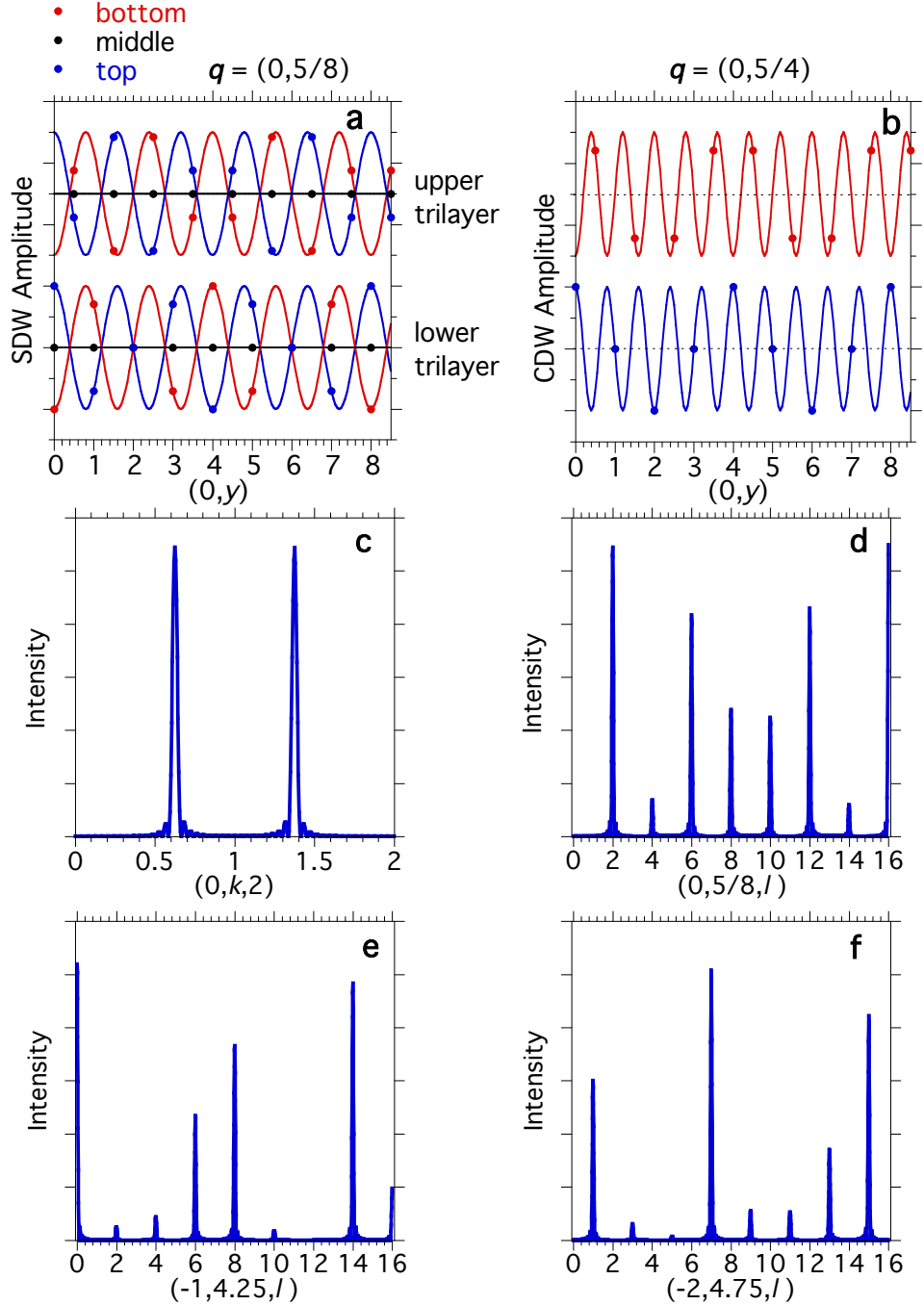

**Fig. 4 | Density wave simulations. a-b**, Model for the SDW and CDW plotted along the $(0,y)$ orthorhombic axis. For **a**, the SDW has a node on the inner plane (middle) and is out-of-phase between the two outer planes (bottom, top). For **b**, the CDW exists on all layers and is in-phase between the layers. In addition, **a** is in-phase between trilayers but **b** is out-of-phase. The shift in the nickel ion positions (dots) in the upper trilayer relative to the lower one is due to the $B$-centered translation. **c-d**, Predicted SDW diffraction intensities along $k$ and along $l$. **e-f**, Predicted CDW diffraction intensities along $l$. **e** and **f** differ in having either even or odd $l$ peaks.

# Supporting Information for:

# Intertwined density waves in a metallic nickelate

Junjie Zhang[1,2]*, D. Phelan[1], A. S. Botana,[3] Yu-Sheng Chen[4], Hong Zheng[1], M. Krogstad[1], Suyin Grass Wang[4], Yiming Qiu[5], J. A. Rodriguez-Rivera[5,6], R. Osborn[1], S. Rosenkranz[1], M. R. Norman[1] and J. F. Mitchell[1]*

[1]Materials Science Division, Argonne National Laboratory, Lemont, Illinois 60439, United States.
[2]Institute of Crystal Materials, Shandong University, Jinan, Shandong 250100, China.
[3]Department of Physics, Arizona State University, Tempe, Arizona 85287, United States.
[4]ChemMatCARS, The University of Chicago, Lemont, Illinois 60439, United States.
[5]NIST Center for Neutron Research, National Institute of Standards and Technology, Gaithersburg, Maryland 20899, United States.
[6]Department of Materials Science, University of Maryland, College Park, Maryland 20742, United States.

*e-mail: junjie@sdu.edu.cn; mitchell@anl.gov

1. Crystal Structure of $La_4Ni_3O_{10}$
2. Determination of CDW propagation vector
3. Correlation length of CDW
4. Neutron scattering in the $hk0$ plane
5. CDW in $Pr_4Ni_3O_{10}$
6. Modeling of the SDW and CDW
7. Real space spin-stripe model calculation
8. Why the CDW is not on the oxygen sites
9. Band structure and Fermi surface nesting

## 1. Crystal Structure of La$_4$Ni$_3$O$_{10}$

The unit cell of La$_4$Ni$_3$O$_{10}$, contains two trilayer perovskite slabs of corner-sharing NiO$_6$ octahedra separated by a rocksalt layer of LaO (see Fig. 2a of main text). Depending on details of the growth conditions and post-growth treatment[1], La$_4$Ni$_3$O$_{10}$ crystallizes in either an orthorhombic (*Bmab*) or monoclinic (*P*2$_1$/*a*) crystal structure above $T_{MMT}$. The crystal studied here shows $T_{MMT}$ = 140 K, indicating that the high temperature structure belongs to the monoclinic space group *P*2$_1$/*a*. However, this structure is pseudo-orthorhombic, and for ease of notation we employ the *Bmab* setting with lattice parameters of $a$~5.41 Å, $b$~5.46 Å, and $c$~27.97 Å throughout this report. In this setting, the two trilayers are symmetry-related by the *B*-centering operation.

## 2. Determination of CDW propagation vector

Linecuts from the 15-ID-D data were used to determine the propagation vector by fitting the SL peaks at (-1, 1.76, 0), (-1, 2.24, 0), (-1, 3.76, 0) and (-1, 4.24, 0) as well as their nearby main Bragg peaks (-1, 2, 0) and (-1, 4, 0) using a Gaussian function. A typical fit is shown in Fig. S1. The obtained values from the Gaussian fit are listed below.

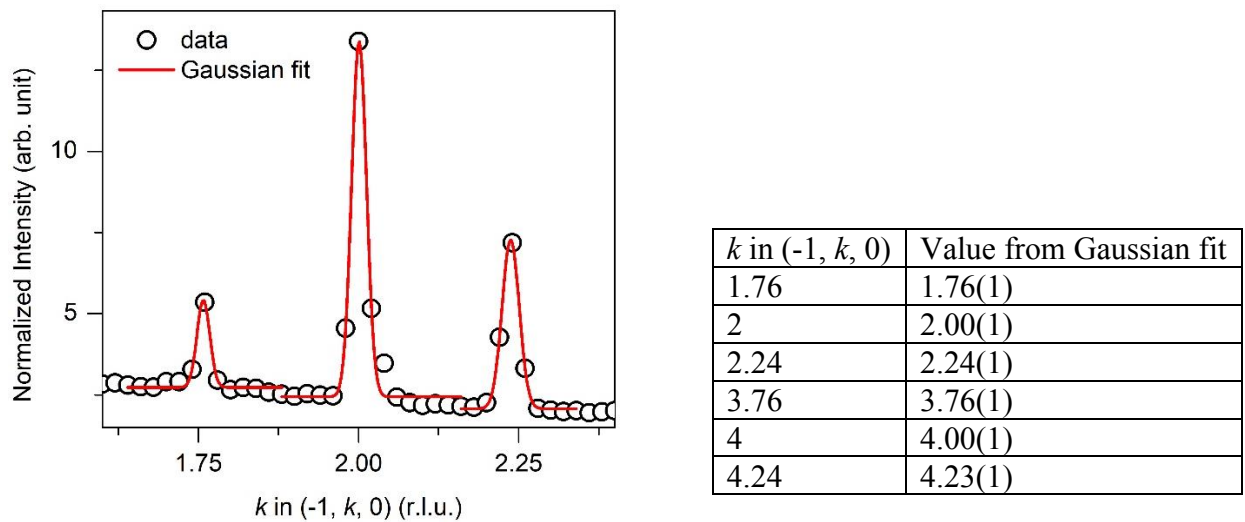

| $k$ in (-1, $k$, 0) | Value from Gaussian fit |
|---|---|
| 1.76 | 1.76(1) |
| 2 | 2.00(1) |
| 2.24 | 2.24(1) |
| 3.76 | 3.76(1) |
| 4 | 4.00(1) |
| 4.24 | 4.23(1) |

**Fig. S1**. Gaussian fit to the main Bragg peak (-1, 2, 0) and SLs.

Considering the limited pixels for each peak from linecuts from 15-ID-D data, we further performed single crystal diffuse scattering experiments at Beamline 33-BM-C at the Advanced Photon Source using a point detector at 70 K ($\lambda$=0.7749 Å). $hkl$ scans were performed using steps of 0.0025 reciprocal lattice unit (r.l.u.) for $k$ scans and 0.01 r.l.u. for $l$ scans. Superlattice peaks were observed and fit to the Gaussian function (see Fig. S2). The obtained wavevector is (2.235-1.753)/2=0.241. The CDW wavevector is calculated from the Bragg peak (130) or (150), thus the wavevector is 0.76$b$*.

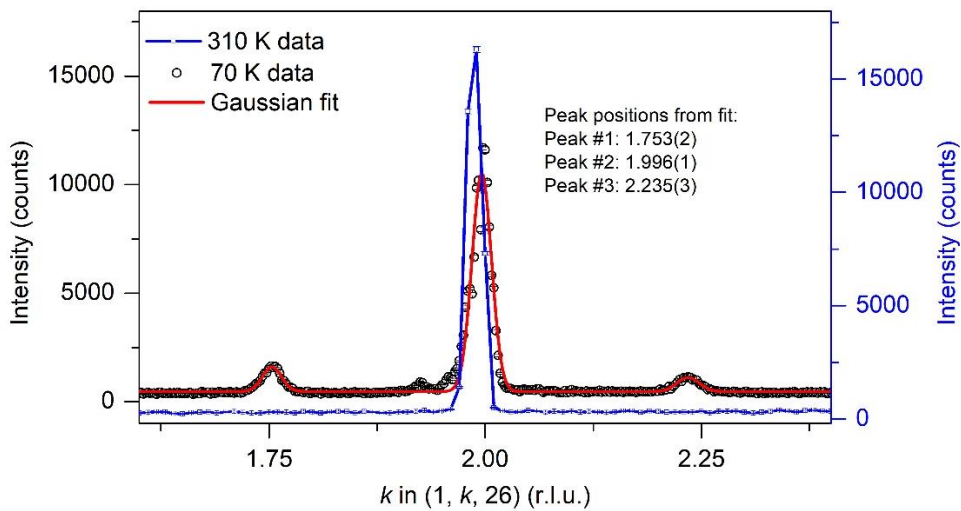

**Fig. S2**. Gaussian fit to the Bragg and SL peaks of $La_4Ni_3O_{10}$.

## 3. Correlation length of CDW

Using data collected at the 33-BM-C beamline of the Advanced Photon Source, the correlation length of the charge order modulation was estimated through analysis of the peak width of the superlattice reflections. The correlation length $\xi$ was determined as $\xi = 1/\Gamma$, where $\Gamma$ is the half-width at half-maximum of the Lorentzian component. A Gaussian fit was performed on the main Bragg peak (1,4,18) in order to estimate the instrumental resolution, and this Gaussian width was fixed in the Voigt function (a convolution of Gaussian and Lorentzian functions) used to fit the superlattice peaks (Fig. S3).

The in-plane and out-of-plane correlation length are $\xi_{ab}$=16-22$b$ (87-118 Å) and $\xi_c$=0.67-0.87$c$ (18.8-24 Å), respectively. The estimated correlation length along $c$ is comparable to the 21 Å that was found independently from 15-ID-D data shown in Fig. S4.

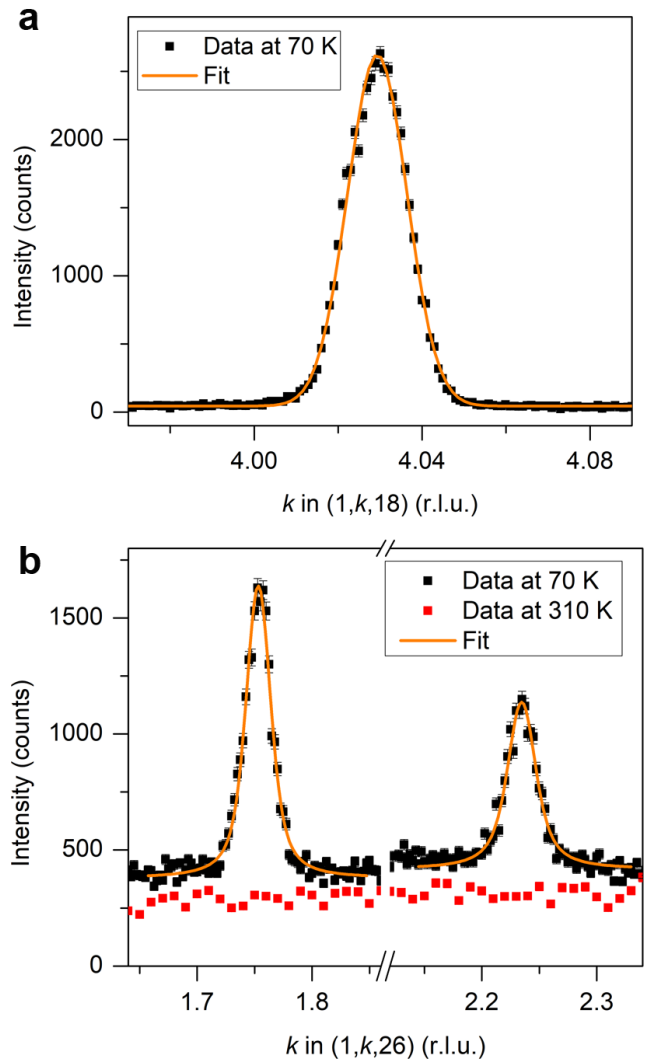

**Fig. S3**. Correlation length of the CDW estimated from data collected at Beamline 33-BM-C at the Advanced Photon Source.

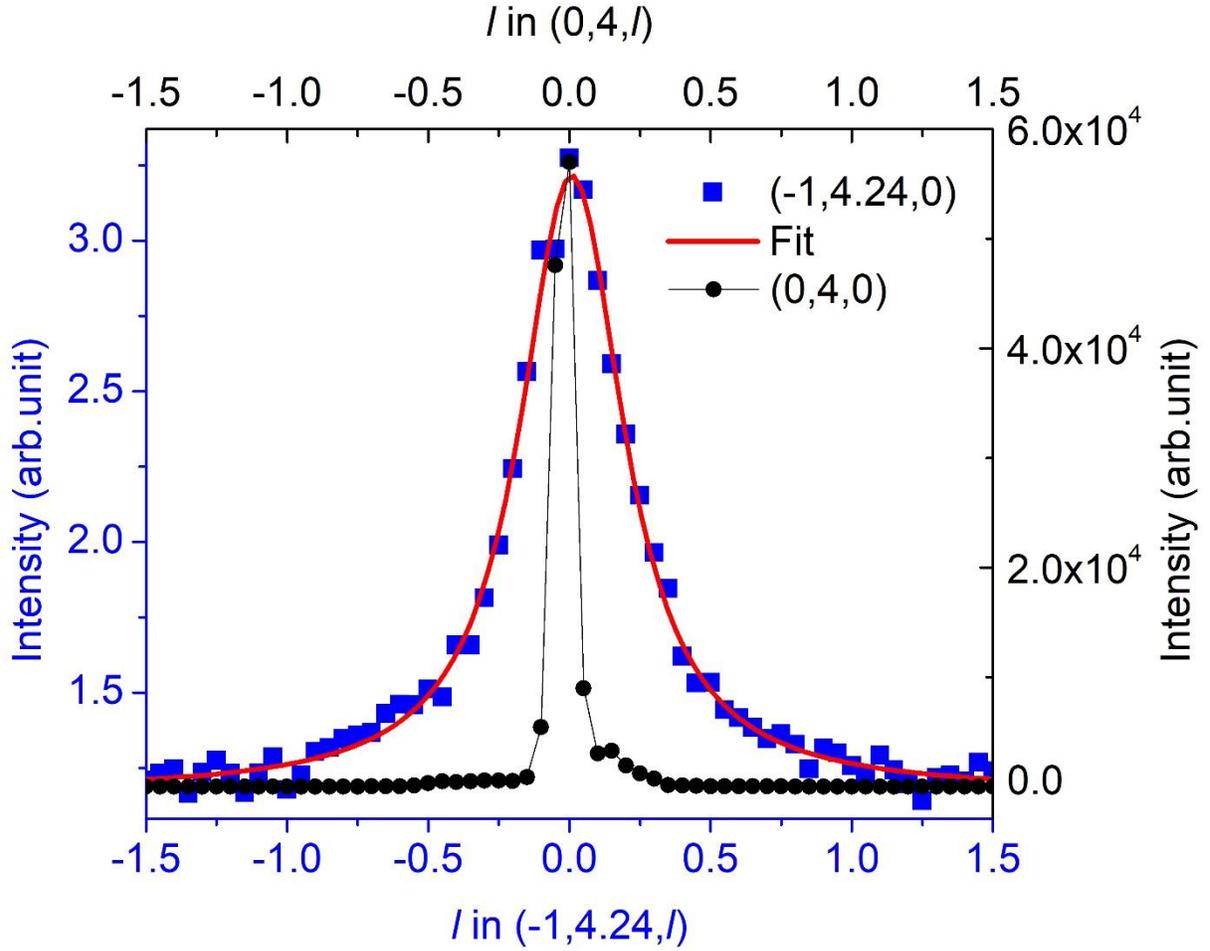

**Fig. S4.** Correlation length of the CDW along the *c* direction (linecuts of 15-ID-D data). The correlation length $\xi$ was determined as $\xi=1/\Gamma$, where $\Gamma$ is the half-width at half-maximum of the Lorentzian component. A Gaussian fit was performed on the main Bragg peak (0,4,0) in order to estimate the instrumental resolution, and this Gaussian width was fixed in the Voigt function (a convolution of Gaussian and Lorentzian functions) used to fit the superlattice peak (-1,4.24,0). Using this procedure, $\Gamma$ was determined to be 0.215 r.l.u. (where 1 r.l.u.=$2\pi/c$) yielding $\xi=0.7c=21$ Å.

## 4. Neutron scattering in the *hk*0 plane

In the main text, neutron scattering measurements were shown in the 0*kl* plane. Here, we show measurements performed in the *hk*0 plane on the same instrument. Please note that whereas the measurements in the 0*kl* plane were performed with one specimen, the measurements in the *hk*0 plane were performed with four co-aligned specimens. A 180 K data set (Fig. S5b) has been subtracted from the 1.6 K data set (Fig. S5a) to reveal the magnetic superlattice peaks from the noise in the raw data. Fig. S5c shows the map with the high temperature subtracted as a background. Fig. S5d shows a line-cut through the temperature subtracted data along *h*. Note that the signal in this plane is much weaker than the maximal signal in the 0*kl* plane because the strong maxima occur at *l*=2 and 6, rather than 0. Nevertheless, a contribution to the *hk*0 plane is observed that may result from intrinsically broad peaks in *l* that are out of plane, exacerbated by the course out-of-plane resolution of the spectrometer.

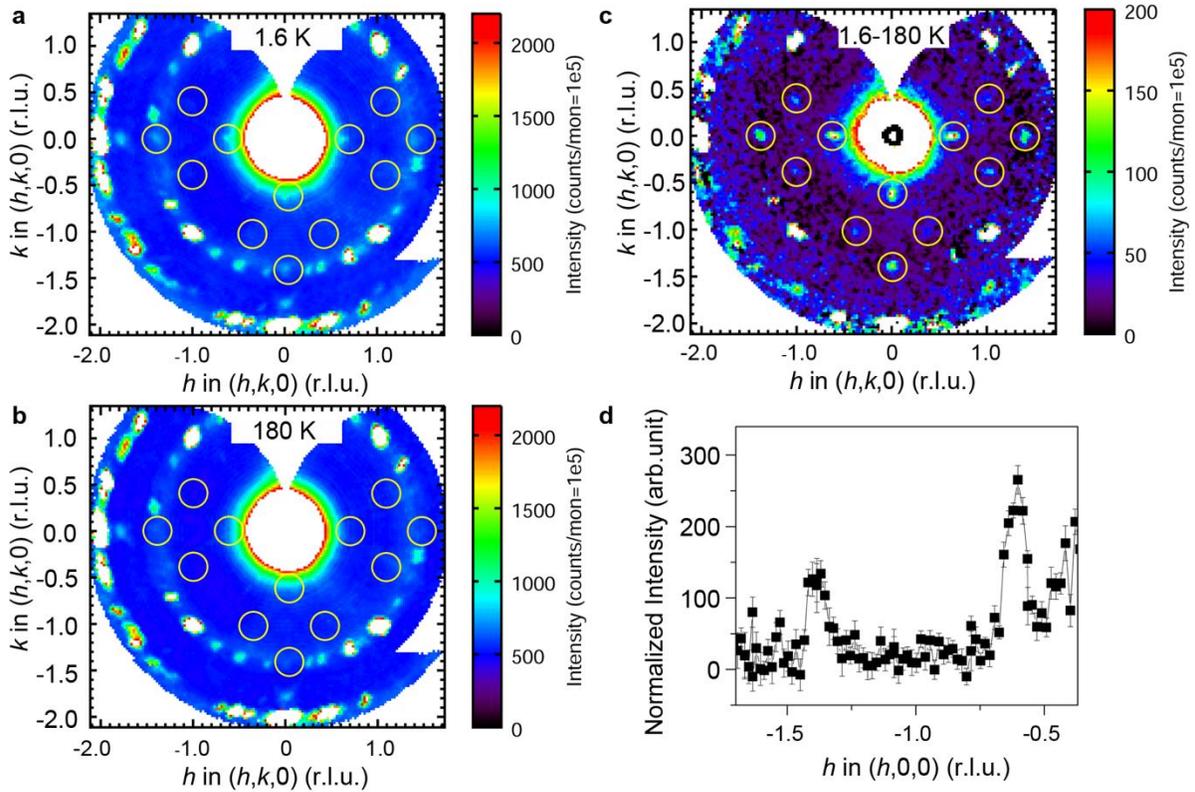

**Fig. S5.** Spin density wave order in $La_4Ni_3O_{10}$. **a** *hk*0 plane at 1.6 K. **b** *hk*0 plane at 180 K. **c** *hk*0 plane with background (180 K) subtracted from the 1.6 K data. **d** Line cuts in **c** along (*h*, 0, 0) showing superlattice reflections at *h*= -0.62 and -1.38. Error bars represent one standard deviation. Open circles indicate the positions of the superlattice peaks.

## 5. CDW in Pr$_4$Ni$_3$O$_{10}$

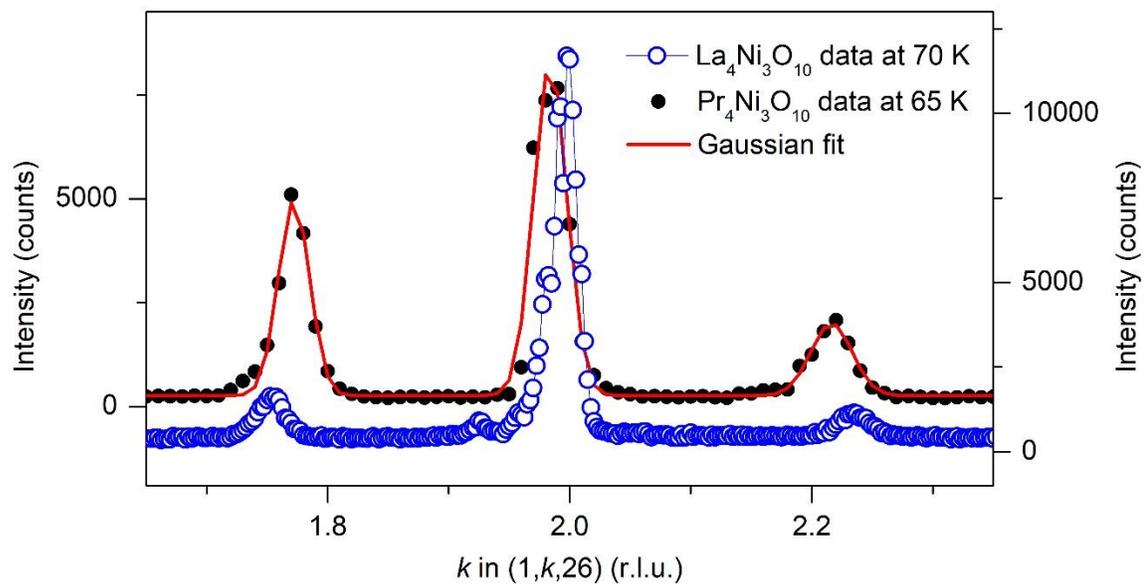

**Fig. S6.** Propagation vector of the CDW in Pr$_4$Ni$_3$O$_{10}$. The peaks were fit using a Gaussian function. The obtained positions are $k$=1.772, 1.984 and 2.216, so the wavevector is 1-(2.216-1.772)/2=0.778 along $b^*$. Note this wavevector (0.78) is slightly different from that of La$_4$Ni$_3$O$_{10}$ (0.76).

## 6. Modeling of the SDW and CDW

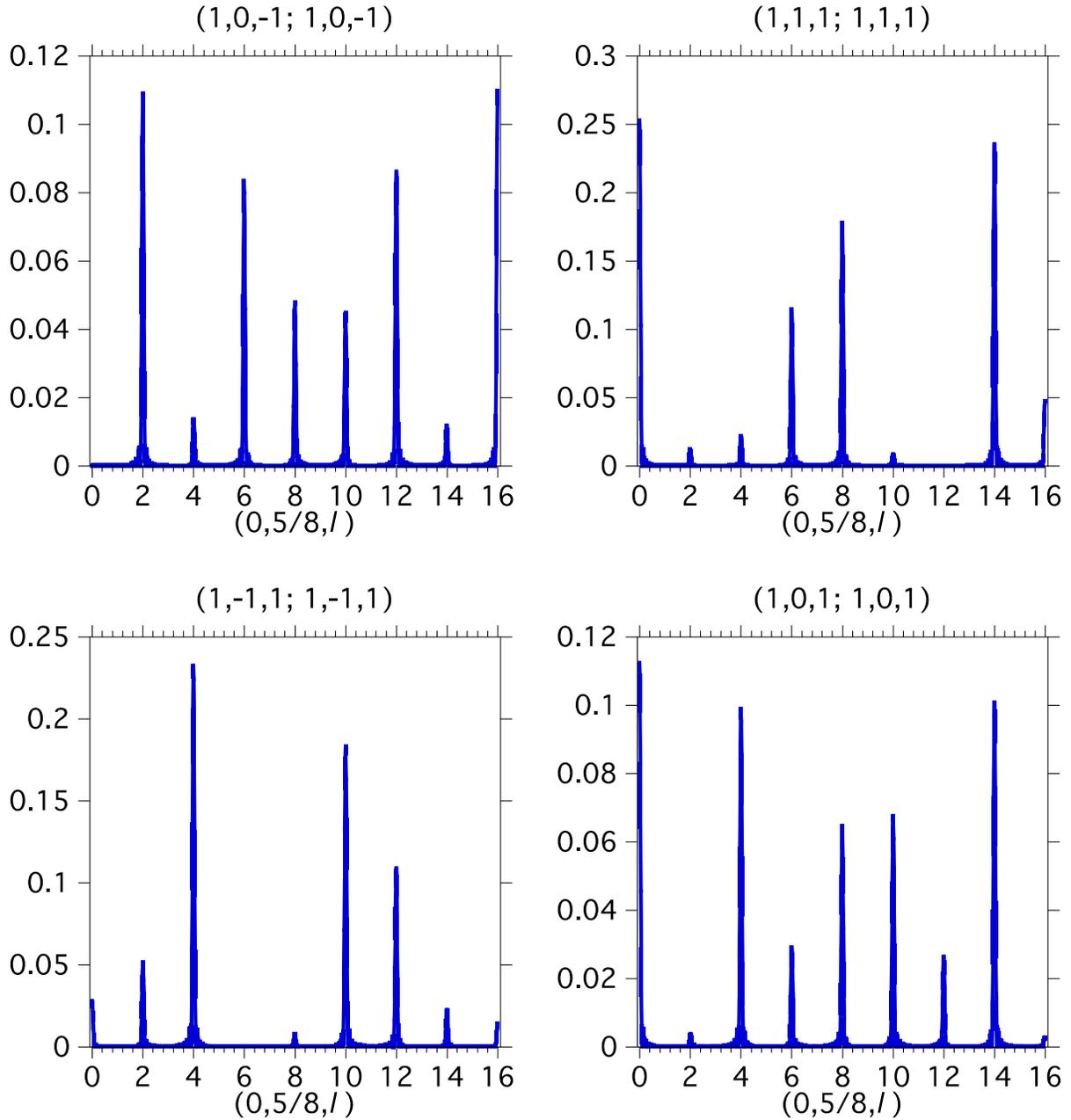

**Fig. S7.** Intensities for different assumed SDW stackings along the *c*-axis for the (0, 5/8, *l*) line cut, calculated using the equation listed in the main text. The titles indicate the stacking pattern along *c* (the $c_z$ weights of the main text). The upper left plot is the SDW model presented in the main text, which has prominent peaks at *l*=2 and 6.

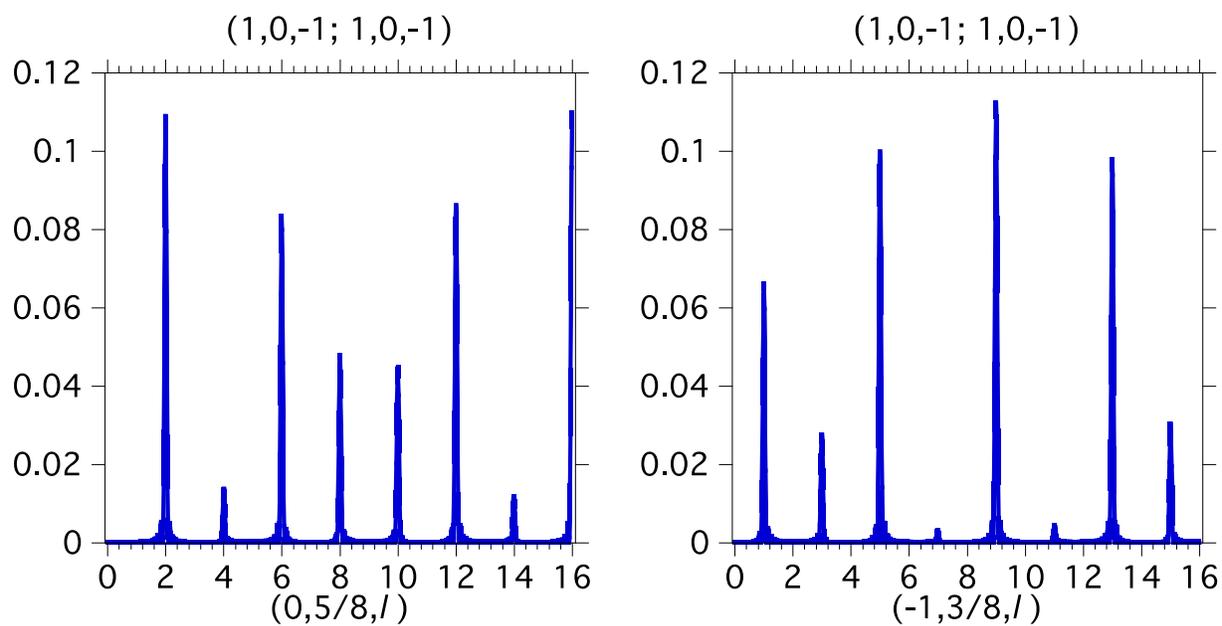

**Fig. S8.** Contrasting the (0, 5/8, *l*) line cut with the (-1, 3/8, *l*) line cut for the SDW model presented in the main text.

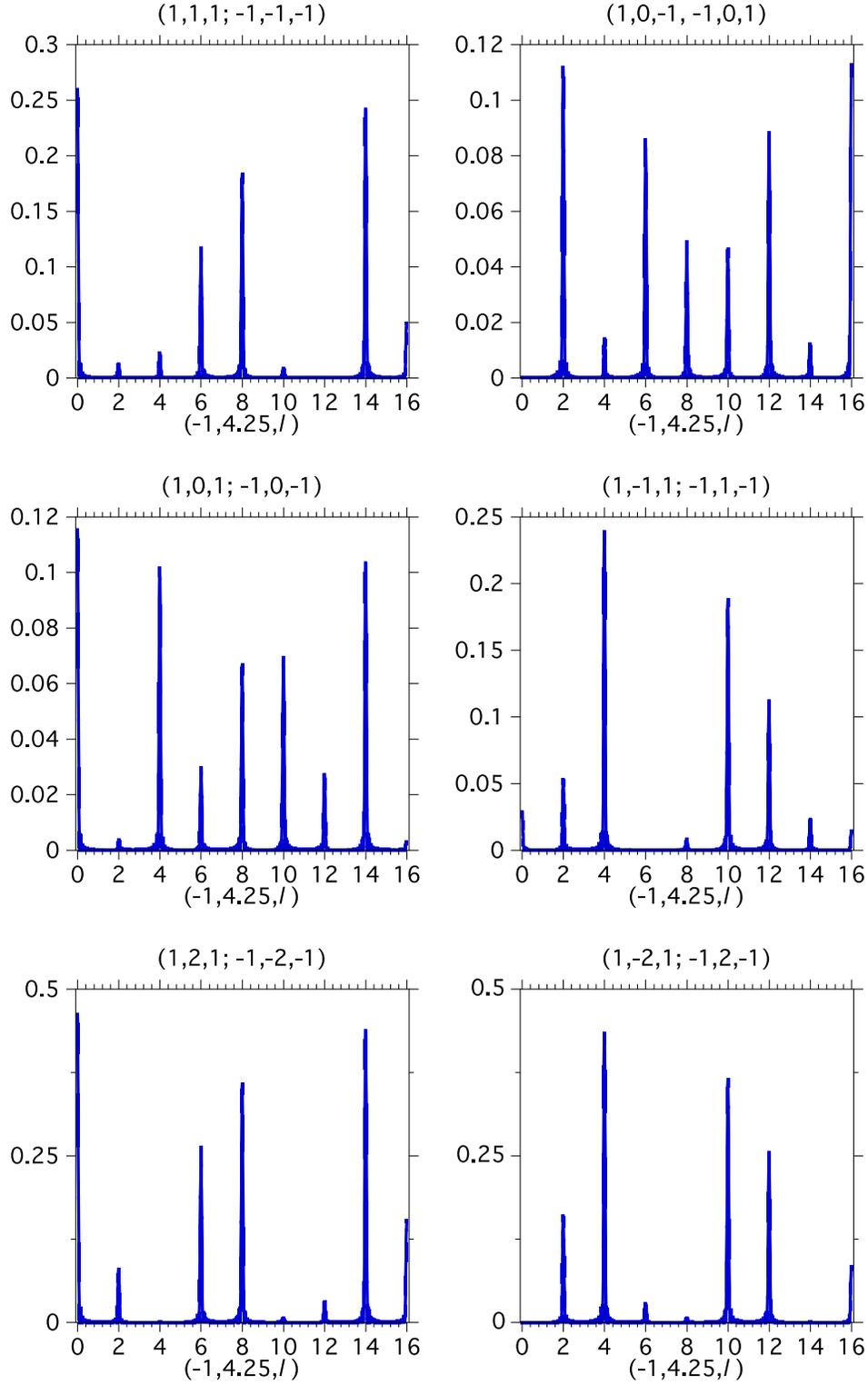

**Fig. S9.** Intensities for different assumed CDW stackings along the *c*-axis for the (-1, 4.25, *l*) line cut. The titles represent the stacking patterns along *c* ($c_z$ weights of the main text). The upper left plot is the CDW model presented in the main text, which has prominent peaks at *l*=0 and 14.

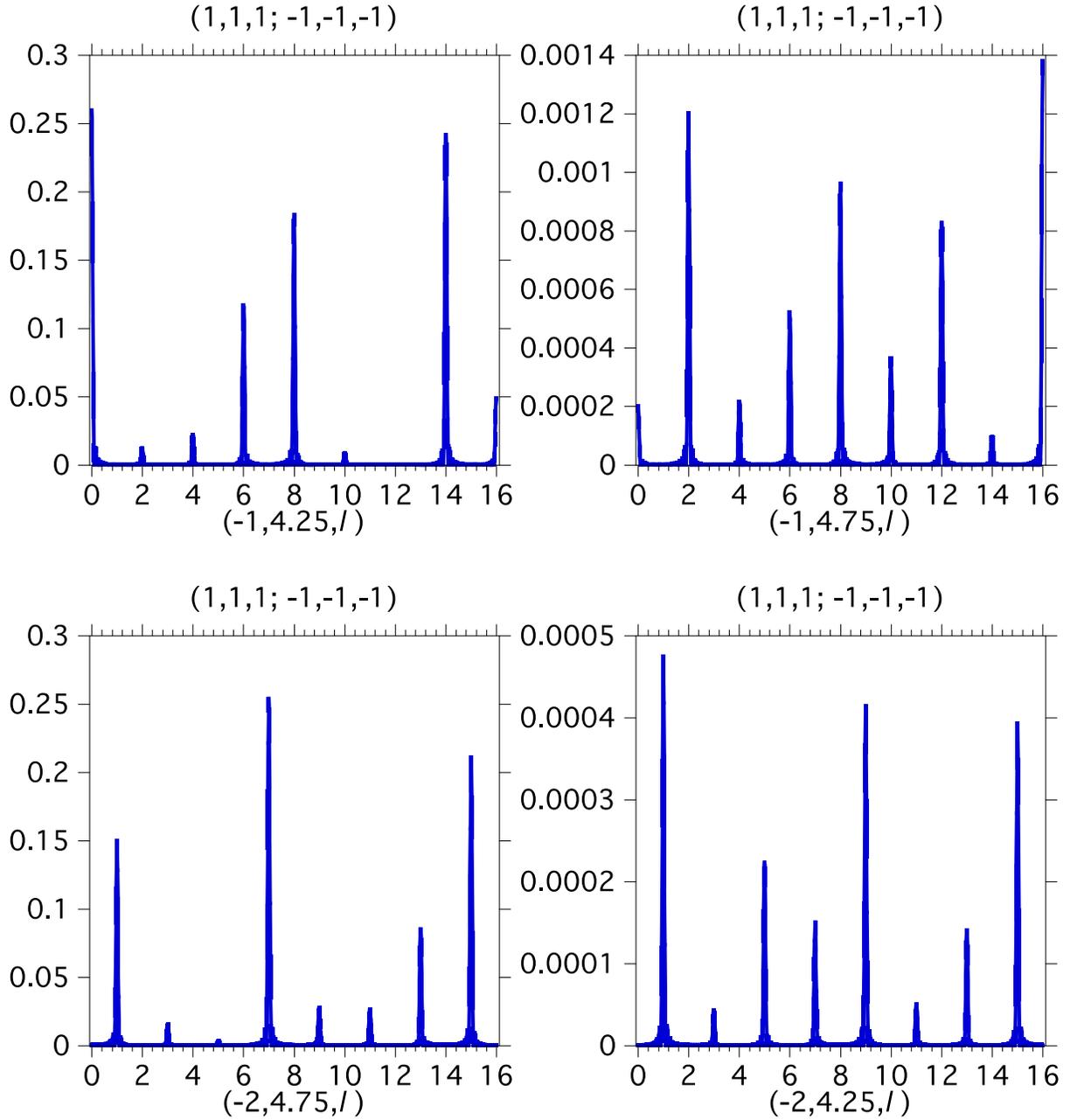

**Fig. S10.** Intensities for various line cuts for the CDW model presented in the main text, contrasting strong CDW peaks (left plots) from weak CDW peaks (right plots). The latter intensities are sensitive to the *Bmab* deviations of the Ni atoms from their *I4/mmm* positions, and so these intensities should be further enhanced in the monoclinic phase (not modeled here).

Above we have presented a simple model that accounts qualitatively for the distribution of intensity in the charge peaks measured by XRD. There are certain systematic discrepancies in the intensity distributions that highlight the need for a more sophisticated model. For example, as shown in Fig. S11, our model for the (-1,k,0) cut of Fig. 2d finds strong peaks at $k\pm q_c$ ($k$ odd) in agreement with the data. For the (0,k,19) cut of Fig. 2e, there are significant differences. In particular, our model indicates strong peaks ($k$ even) and weak peaks ($k$ odd), whereas the data find these peaks to be of comparable intensity. Moreover, the SL intensities are reduced relative to those in Fig. 2d by $10^2$ as compared to about $10^1$ found in experiment. To successfully reproduce the data, one would need a more sophisticated model than the present one. For instance, taking into account the monoclinic distortion is non-trivial in the model used here since $\beta \neq 90$ degrees, meaning the simple cosine function would be de-phased when going from one layer to the next. Moreover, for hard x-rays, one is most sensitive to the atomic displacements. A strain wave model based on displacements of La, Ni, and O ions is a natural next step, albeit an involved endeavor, as can be seen from studies in the cuprates[2].

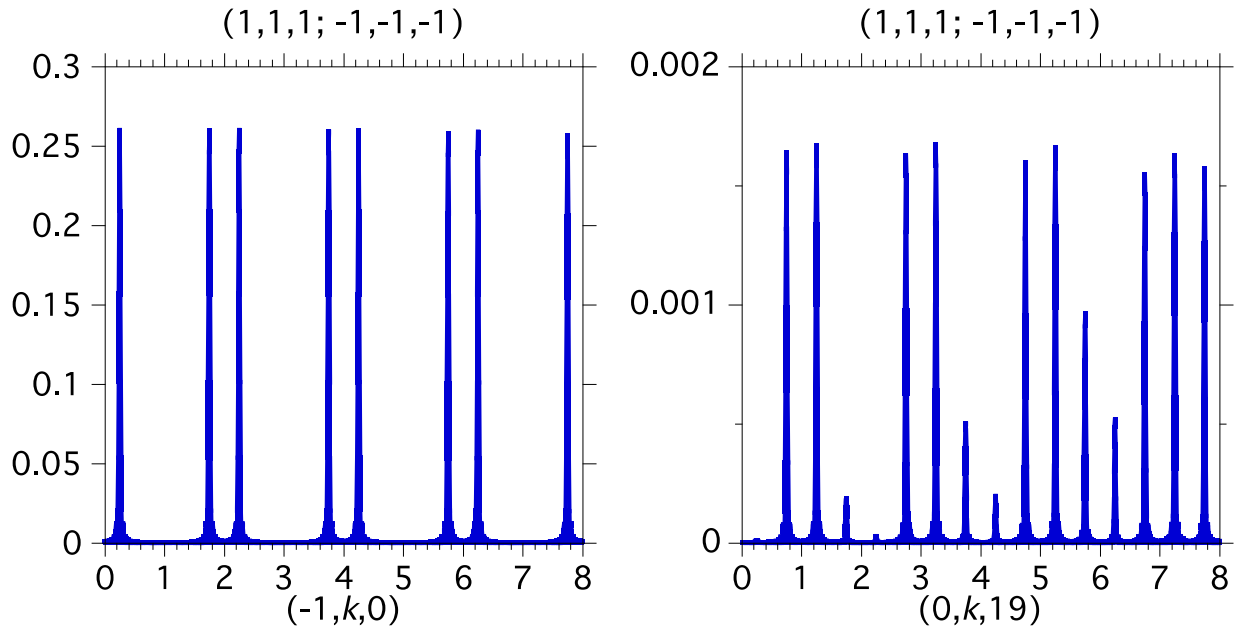

**Fig. S11**. Intensities from the CDW model presented in the main text for the two $k$ cuts shown in Fig. 2. For the right plot, note the alternation of strong peaks ($k$ even) with weak peaks ($k$ odd), in contrast to the data where these peaks have comparable intensity.

## 7. Real space spin-stripe model calculation

Various real space models for the SDW can be constructed by assuming 16 Ni ions per plane per magnetic unit cell. With an appropriate pattern of spins and holes, the primary peak at $1-q_s=5/8$ can be generated, but because of the real space (square wave) form, (odd) harmonics appear with the strongest at $2-3q_s=7/8$, albeit with an intensity reduced by an order of magnitude relative to $1-q_s=5/8$ (Fig. S12). Given the absence of these harmonics in the data and the fact that the real space pattern does not correspond to the correct doping relative to $LaNiO_3$ (which would imply a $1-q_s=2/3$ instead), we did not pursue real space models further.

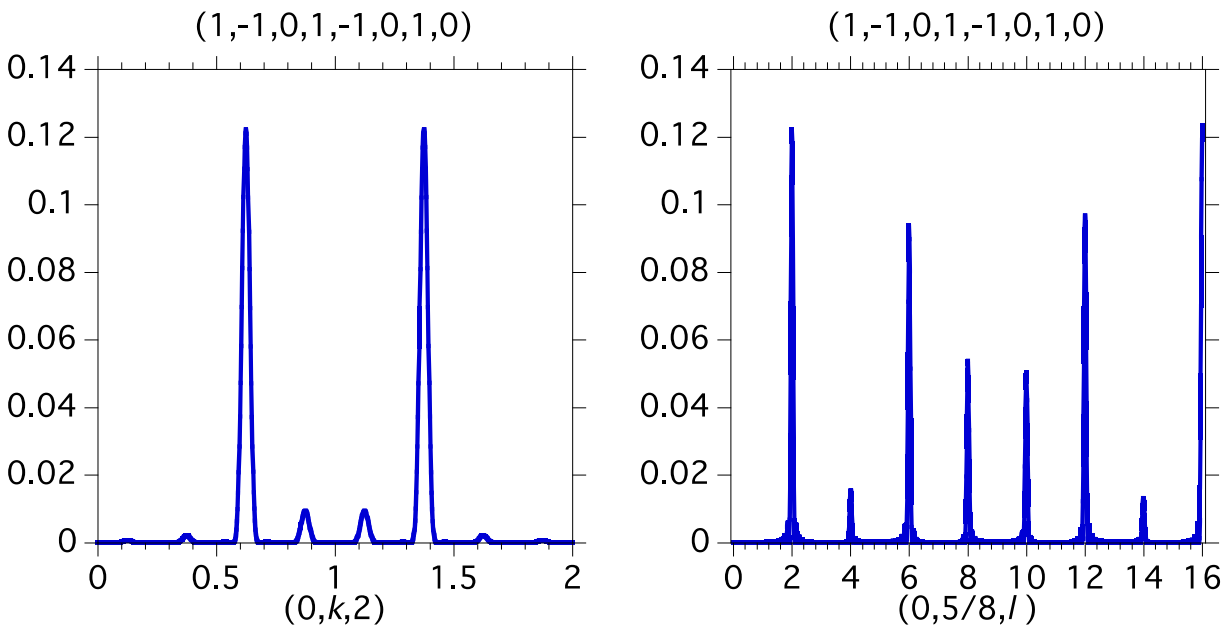

**Fig. S12**. Intensities from a real space stripe model. The title of the plots shows the SDW weights on each Ni ion for eight successive diagonal rows. For the eight rows beyond this, the spins are reversed, with the pattern repeated. The stacking along $c$ is the same as before ($\uparrow, -, \downarrow$), with the lower and upper trilayers of a given unit cell being in phase. This acts to maintain peaks at even $l$. Since there are five non-zero weights for each eight rows, this gives rise to a primary peak at $1-q_s=5/8$. The secondary peaks ($k=7/8$, etc.) are odd harmonics due to the "square wave" nature of the stripe, as compared to the sinusoidal behavior of the itinerant model.

## 8. Why the CDW is not on the oxygen sites

Our model assumes the density wave is centered on the Ni sites. This is sensible for the SDW, but what about for the CDW? A simple argument shows that the CDW cannot involve the planar oxygen sites. The in-plane planar oxygen coordinates (*Bmab*) are (¼,¼), (¾,¼), (¼,¾), and (¾,¾). For the first pair at $y=1/4$, the CDW will have the same amplitude, and similarly for the second pair at $y=3/4$. Therefore, for *h* odd, the sum over a pair vanishes because of the $e^{i\mathbf{k}\cdot\mathbf{r}}$ structure factor, since their *x* coordinates differ by ½. This is in clear contradiction to experiment. If instead one assumes a *d*-wave form factor for the CDW, where the CDW amplitude has opposite phase on the two members of the pair, then the sum vanishes for *h* even, again inconsistent with the data. We argue that the small monoclinic distortion found in $La_4Ni_3O_{10}$[1] can only cause weak violations of these rules. Finally, we remark that high energy x-rays are primarily sensitive to atomic displacements. Not only would such a strain wave differ from the CDW itself, but also other ions, such as the apical oxygens and La, can contribute, as seen in cuprates[2]. More information than we have at present would be needed to model such a complicated strain wave.

## 9. Band structure and Fermi surface nesting

Given the above context, we can ask whether Fermi surface nesting is relevant to the physics of $La_4Ni_3O_{10}$, as demonstrated in chromium[3] and suggested as well for $LaNiO_3$[4] (recognizing that ARPES has been agnostic on this point for $La_4Ni_3O_{10}$[5]). We note that Seo et al.[6] have proposed from extended Hückel tight-binding calculations that the MMT arises from a CDW instability resulting from Fermi surface nesting. To explore this possibility further, we performed DFT calculations for $La_4Ni_3O_{10}$ using *Bmab* coordinates. We then fit the eight bands nearest to the Fermi energy with a Fourier series spline fit. The fit is shown in Fig. S13a and the resulting Fermi surfaces in the $k_z$=0 plane in Fig. S13b. Note the zone folding relative to a hypothetical undistorted *I4/mmm* structure. Since the bands 1 and 2 Fermi surfaces are hole-like, and the bands 3 and 4 Fermi surfaces are electron-like, nesting is a distinct possibility. To test this possibility, we first accurately determined the Fermi energy and then calculated the static susceptibility ($\chi$) using a linear tetrahedron method. The resulting interband $\chi$ terms are shown in Fig. S13c. Here we present a cut along ($q$,0,0). Almost equivalent results are found along (0,$q$,0), implying that the orientation of the density wave with ***q*** ∥ ***b*** is due to pinning by the structure. The band 2 to band 3 interband term shows a maximum at (0.5,0,0) but has subsidiary peaks at (0.5±0.11,0,0), with 0.61 lying close to the observed 1-$q_s$=0.62.

Since we find considerable structure in the density of states near the Fermi energy, we recalculated $\chi$ by reducing the Fermi energy by 30 meV (Fig. S13d). Now the side peaks become the global maxima, although they are shifted to (0.5±0.09,0,0). These results indicate that nesting is a distinct possibility. To improve on these results would require more sophisticated calculations that include not only spin matrix elements between the various band eigenvectors, but also potential correlation-induced energy shifts between the Ni $3d$ and O $2p$ states. For these reasons, we did not pursue further calculations along these lines, including the $q_z$ dependence (which we find to be weak).

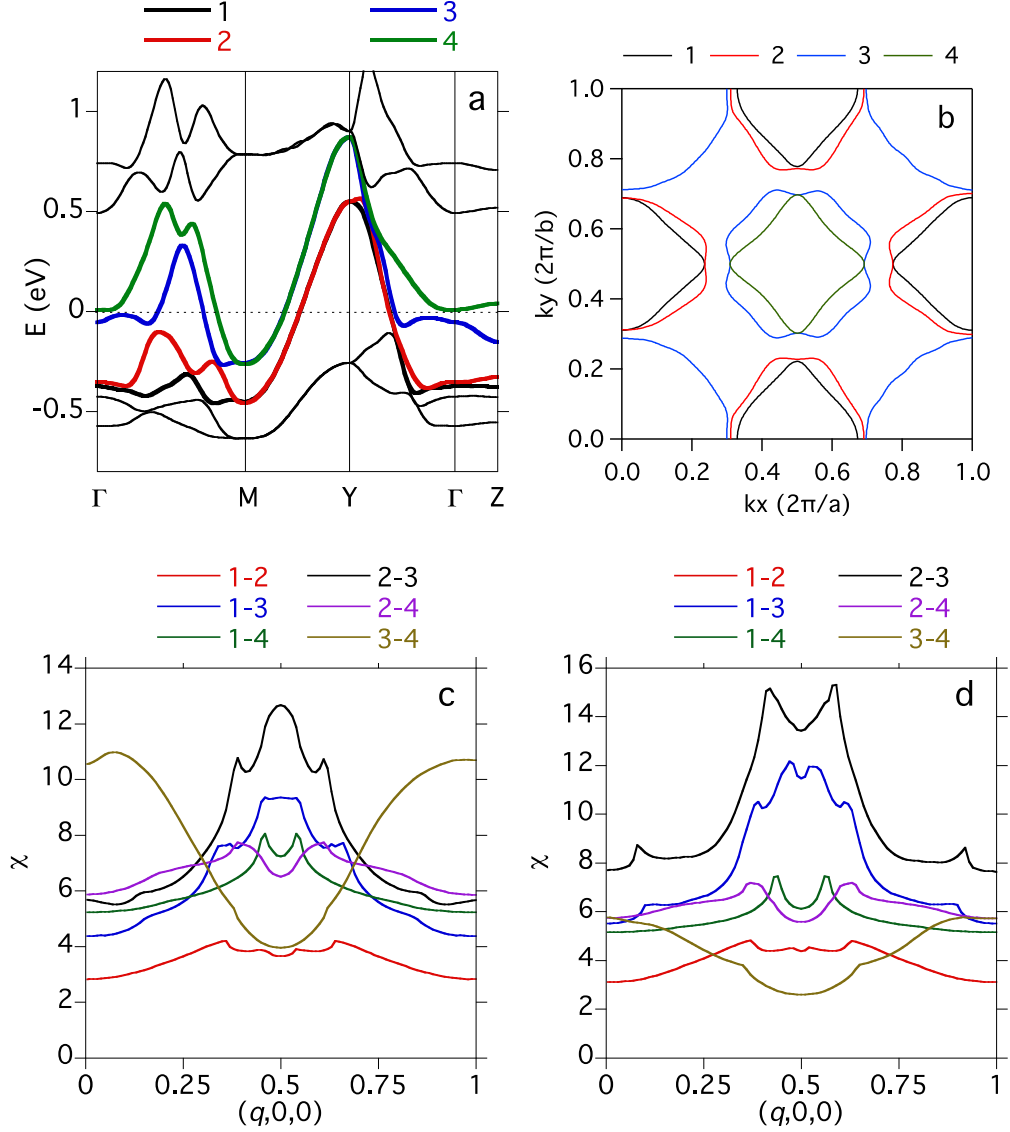

**Fig. S13.** Band structure and Fermi surface nesting. **a**, Fourier series spline fit to the eight DFT bands nearest the Fermi energy (energies in eV with Fermi energy at 0). Here, $M=(½,½,0)$, $Y=(0,½,0)$, and $Z=(0,0,1)$ in *Bmab* notation ($2\pi/a$, $2\pi/b$, $2\pi/c$ units). The bands used in (b-d) are marked in bold and labeled. **b**, Fermi contours ($k_z=0$ plane) for the four energy bands crossing the Fermi energy, with bands 1,2 being hole-like and 3,4 being electron-like, implying the possibility of nesting. Decomposition of the susceptibility χ into various interband terms (ij, where i and j are band indices, with each being the sum ij+ji) at **c** the Fermi energy and **d** at 30 meV below the Fermi energy. Here, $q$ is in $2\pi/a$ units. The outermost of the three peaks in **c** for the 2-3 interband term is at $q=0.61$, close to the observed SDW wavevector at $1-q_s=0.62$. This peak becomes a global maximum in **d**, though it is shifted to $q=0.59$. This indicates that nesting is a potential cause of the SDW in $La_4Ni_3O_{10}$.